\documentclass[apjl]{emulateapj}
\usepackage{epstopdf}
\usepackage{natbib}
\usepackage{amsmath}
\usepackage{enumerate}
\begin{document}

\title{Electromagnetic Counterparts to Black Hole Mergers Detected by LIGO}
\author{Abraham Loeb\altaffilmark{1}}
\altaffiltext{1}{Department of Astronomy, Harvard University, 60 Garden St., Cambridge, MA 02138}

\email{E-mail: aloeb@cfa.harvard.edu}

\begin{abstract}
Mergers of stellar-mass black holes (BHs), such as GW150914 observed
by LIGO, are not expected to have electromagnetic counterparts.
However, the Fermi GBM detector identified a $\gamma$-ray transient
0.4 s after the gravitational wave (GW) signal GW150914 with
consistent sky localization. I show that the two signals might be
related if the BH binary detected by LIGO originated from two clumps
in a dumbbell configuration that formed when the core of a rapidly
rotating massive star collapsed. In that case, the BH binary merger
was followed by a $\gamma$-ray burst (GRB) from a jet that originated
in the accretion flow around the remnant BH. A future detection of a
GRB afterglow could be used to determine the redshift and precise
localization of the source. A population of standard GW sirens with
GRB redshifts would provide a new approach for precise measurements of
cosmological distances as a function of redshift.

\end{abstract}

\section{Introduction}\label{s:intro}
The detection of the gravitational wave (GW) source GW150914 by the
Laser Interferometer Gravitational Wave Observatory (LIGO) was
interpreted as the merger of a black hole (BH) binary whose members
have masses of $M_1=36^{+5}_{-4}M_\odot$ and $M_2=29^{+4}_{-4}M_\odot$
\citep{Abbott}. The GW signal exceeded the background noise level of
LIGO for the last $\sim 0.2$ s of the merger when the BH binary
separation was shorter than $\sim 10GM/c^2$, where $M=(M_1+M_2)$. A
merger of two BHs in vacuum is expected to have no electromagnetic
counterpart. But nature is sometimes more imaginative than we are.

The Gamma-ray Burst Monitor (GBM) on board the {\it Fermi} satellite
reported the detection of a transient signal at photon energies $>50$
keV that lasted 1 s and appeared 0.4 s after the GW signal
\citep{Fermi16}. The GBM signal encompasses 75\% of the probability
map associated with the LIGO event localization on the sky.

Below we explore the possibility that the GW and Gamma-Ray Burst (GRB)
signals originated from a common origin, namely a single,
rapidly-rotating, massive star.\footnote{The GRB luminosity requires a
  mass accretion rate that exceeds the Eddington limit by more than 9
  orders of magnitude. Such high infall rates are naturally supplied
  during the collapse of a massive star. Alternative scenarios in
  which a neutron star joins a BH binary during its final merger phase
  or a pre-existing BH sinks to the center of a massive star just
  around the time when the core of the star collapses to make the
  second BH \citep{Janiuk} and produce a GRB merely a fraction of a
  second after the merger, require more fine-tuning in the initial
  conditions of the system.}  As the core of the star collapsed, it
broke into two clumps in a dumbbell configuration.  The two clumps
collapsed separately into two BHs which eventually merged due to GW
emission. The GRB was produced from an outflow generated by the
merging BHs or from a jet emanating out of the accretion disk of
residual debris around the BH remnant, similarly to the collapsar
model of long-duration GRBs \citep{Mac,Woo}. The mass accreted during
the inspiral must have been a small fraction of $M$ given the good
match between the observed LIGO signal and the theoretical GW template
for a BH binary in vacuum. The low accretion rate during the inspiral
is naturally explained by the clearing of a central cavity that is
expected for a circumbinary disk around a binary BH system
\citep{Hayasaki,Cuadra,Colpi,Kocsis,Farris}.

The Fermi-GBM detection was not reproduced by the INTEGRAL satellite
\citep{INT}. The a posteriori nature of the GW150914-GBM detection
raises additional concerns about its reality. With many more LIGO
events expected in the future, it would be straightforward to test
whether GRBs are a common by-product of BH-BH mergers. The mechanism
considered in this {\it Letter} offers motivation for conducting a
systematic search for GRB counterparts to all LIGO sources.

\section{Core Collapse into a Black Hole Binary}
The prevailing collapsar paradigm for long-duration GRBs involves the
collapse of the core of a massive star to a single BH \citep{Woo}.  In
order to produce a GRB outflow, the infalling matter must have a
sufficiently high specific angular momentum, $j \gtrsim 3\times
10^{16}~{\rm cm^2 s^{-1}}$ \citep{Mac}, so that its centrifugal
barrier lies outside the innermost stable circular orbit (ISCO) around
the BH.

To explain the coincidence between a GRB and GW150914 as well as the
full temporal window during which LIGO detected a GW signal, we
hypothesize that a BH binary formed during the collapse of a rapidly
rotating star with an initial orbital radius of $R_b\gtrsim 10GM/c^2
\sim 10^8~{\rm cm}$ (corresponding to a binary separation of $2R_b$
for $M_1\sim M_2$). The centrifugal barrier of the infalling matter is
outside this radius as long as,
\begin{equation}
j = \left({GM R_b}\right)^{1/2} \gtrsim \sqrt{10} {GM\over c}\sim
10^{18}~{\rm cm^2~s^{-1}} .
\label{eq:1}
\end{equation}

Given that the core of the star needs to be more massive than $M=
65^{+9}_{-8}M_\odot$, the progenitor must be a massive star with a
total mass that exceeds $100 M_\odot$. A natural path to making such a
star would be the merger of two less massive stars that are born in a
tight binary. Each of the parent stars could have had a helium core
with a mass below $35M_\odot$, avoiding the pulsational pair
instability that is capable of dispersing the core
\citep{Heger,WooB,WH}. Once the two stars merge, the merger product
will be endowed with rapid rotation.

There is strong evidence that single massive stars often originate
from the merger of two lower-mass stars \citep{deMink1,deMink2}. The
stellar evolution of the merger product is significantly different
from the standard evolution of an isolated star due to the chemical
mixing and rapid rotation that result from the merger
\citep{deMink1,Hwang,Mandel}. The channel envisioned here for the
birth of BH binaries within the core of a single massive star could be
realized in only a small fraction of all massive stars and still be
within the wide range of BH-BH merger rates that are consistent with
the detection of GW150914 \citep{LIGO2}.


Very massive stars of mass $M_\star\gtrsim 100M_\odot$ are dominated
by radiation pressure and hence their luminosity is close to the
Eddington limit \citep{Bond,Bromm,LF},
\begin{equation}
L_E=1.3\times 10^{40}\times \left({M_\star\over
    100M_\odot}\right)~~{\rm erg~s^{-1}}.
\end{equation}
Since their effective surface temperature, $T_s\sim 10^5~{\rm K}$, has
only a weak dependence on mass \citep{Bromm}, their radii are
approximately given by \citep{LF},
\begin{equation}
R_\star=\left(L_E\over 4\pi \sigma T_s^4\right)^{1/2}\approx 4.3\times
10^{11} \left({M_\star\over 100M_\odot}\right)^{1/2}~{\rm cm} ,
\end{equation}
where $\sigma$ is the Stepfan-Boltzmann constant. To remain
gravitationally bound, the stars must have a specific angular momentum
that is significantly lower than
\begin{equation}
j_{\rm max}= \left({GM_\star R_\star}\right)^{1/2}=7.6\times
10^{19}\left({M_\star\over 100M_\odot}\right)^{3/4}~{\rm cm^2s^{-1}}.
\end{equation}

Assuming hydrostatic equilibrium and electron scattering opacity, one
can show that very massive stars are convectively unstable \citep[see
  Appendix of][]{LR}. With elastic isotropic scattering of the
convective blobs, the star would admit solid body rotation
\citep[although differential rotation is expected for more realistic
  cases; see][]{Kumar}. For a fixed rotation frequency $\Omega$, the
specific angular momentum would have the profile $j=\Omega r^2$, with
$r$ being the cylindrical radius from the rotation axis. The
constraint in equation (\ref{eq:1}) can therefore be rewritten as
\begin{equation}
{j_s\over j_{\rm max}} \gtrsim 1.3\times 10^{-2}\left({R_c\over
  R_\star}\right)^{-2} \left({M_\star\over 100M_\odot}\right)^{-3/4},
\end{equation}
where $j_s\equiv \Omega R_\star^2$ and $R_c\gtrsim 0.1R_\star \gg
R_b\sim 10^8~{\rm cm}$ is the radius of the core that collapses to
make the BH binary. We therefore conclude that the progenitor star
must have been rapidly rotating, not much below its break-up
frequency.  

A rapidly rotating progenitor would be the natural outcome of the
merger between two stars in a tight binary system with a common
envelope. As discussed above, the merger of a binary star system is a
common channel for producing a progenitor star of the needed mass to
explain GW150914 \citep{deMink1,deMink2}. Ejection of the hydrogen
envelope during the merger would be needed, since a red supergiant
envelope would choke the BH jet and suppress the appearance of a short
GRB. In addition, the restriction on a weak mass loss through a
stellar wind (to maintain a high progenitor mass) during nuclear
burning would favor a progenitor of low metallicity.  The evolution of
this progenitor star would be non-standard due to its rapid rotation
and anomalous chemical composition and stratification after its
evolutionary clock was reset by the merger, similarly to blue
stragglers \citep{Sills}. The star would evolve by burning hydrogen
into heavier elements up to iron, and eventually develop a layered
core structure that loses pressure support and collapses. During the
burning stages, the central region of the star would contract, spin
more rapidly, and develop strong differential rotation.

In our model, the BH binary forms out of the collapse of a rapidly
rotating helium core of more than $\sim 65M_\odot$ (but less than
twice this value to avoid pair instability), which surrounds an iron
core of more than $\sim 5M_\odot$ in hydrostatic equilibrium before
the collapse. Furthermore, rapid rotation is needed to stabilize the
core against an explosion. The specific angular momentum of the iron
core needs to exceed a few times $10^{17}~{\rm cm^2~s^{-1}}$ in order
for it to fission into two clumps. After its collapse, the iron core
would form a flattened, rapidly-rotating configuration that cools
through neutrino emission. The resulting disk-like structure is
unstable to the formation of a bar that breaks into two clumps. Each
clump collapses to a BH and the BHs grow in mass by accreting most of
the surrounding carbon and oxygen core within a free fall time of
about a minute after their formation. If the two BHs achieve their
final masses of $\sim 30M_\odot$ at a separation $a$, their subsequent
merger time due to the emission of GWs would be $t_{\rm GW}\sim 4~{\rm
  min}\times (a/5\times 10^8~{\rm cm})^4$. Additional accretion of
core material onto the remnant BH would lead to the formation of the
GRB jet. The strong dependence of the GW merging time on clump
separation implies that only a subset of all rapidly-rotating massive
stars might have the conditions that lead to the birth of a $\sim
(30M_\odot+30M_\odot)$ BH binary followed by a GRB jet, as in
GW150914-GBM. Many more cases may lead to a GRB without a GW signal or
to a GW signal with a choked GRB. The deposition of jet energy in the
envelope of the star will likely lead to a supernova explosion.

The appearance of a dumbbell configuration in a collapsing, rapidly
rotating system was considered in the literature as a path towards the
formation of common envelope massive star binaries through fission
\citep{Tohline,NewT} as well as binaries of supermassive BHs from the
collapse of supermassive stars \citep{Ott}.  In particular, the
general relativistic simulation of \cite{Ott} demonstrated that a
rapidly differentially-rotating star without nuclear burning could
produce a bar that breaks into two clumps of comparable masses,
consistently with the similarity between $M_1$ and $M_2$ in GW150914.
Efficient neutrino cooling or magnetohydrodynamic processes are
required to enable rapid collapse of each clump to a BH
\citep{Narayan,Liu}.  The formation of a disk \citep{Fryer} may
represent an intermediate step before the bar instability and clump
formation identified by the simulations of \cite{Ott}.

An alternative path to forming a BH binary inside the envelope of a
massive star would involve maintaining the identity of the two helium
cores of the merging progenitor stars as they orbit each other and
collapse separately into two BHs surrounded by a common envelope. As
the orbit of the resulting binary BH shrinks due to GW emission,
residual matetial may accrete to make the GRB jet. However, it is
unclear whether the highly super-Eddington accretion rate required by
a GRB can be achieved in this case.

The LIGO limits on the spin amplitude of the two BHs are rather weak
($a_1<0.69\pm 0.05$ and $a_2<0.88\pm 0.10$). The final spin of the
remnant BH inferred by LIGO is $0.67^{+0.05}_{-0.07}$, but the
subsequent accretion of matter could endow it with additional spin and
promote the production of a GRB outflow.

A massive BH binary is expected to eventually clear a central cavity
of twice its semi-major axis in the surrounding circumbinary disk
\citep{Hayasaki,Cuadra,Colpi,Kocsis,Farris}. The delay in filling up
this cavity after the BHs' final plunge inside the ISCO would be of
order the ISCO dynamical time, which is much shorter than the 0.4 s
delay between the GRB and GW150914. For a progenitor star in the mass
range $M_\star=10^{2}$--$10^3M_\odot$, most of the observed 0.4 s
delay can be accounted for by the neutrino cooling timescale or by the
extra time it takes the GRB jet to cross the star relative to GWs for
a jet Lorentz factor in the range $\gamma\sim 4$--$7$.

\section{Discussion}
We described a novel mechanism for a prompt electromagnetic
counterpart to the merger of stellar-mass BH binaries, such as
GW150914.  The proposal was motivated by the Fermi GBM detection of a
$\gamma$-ray transient 0.4 s after GW150914 \citep{Fermi16}. Even if
these two signals are unrelated, the possible existence of
electromagnetic counterparts to BH mergers at cosmological distances
argues in favor of sending LIGO alerts for follow-up observations by
radio, infrared, optical, UV, X-ray and $\gamma$-ray telescopes.

The inferred GRB luminosity for GW150914-GBM (at photon energies
between 1 keV and 10 MeV) of 1.8$^{+1.5}_{-1.0}\times 10^{49}~{\rm
  erg~s^{-1}}$ and its measured duration of 1 s \citep{Fermi16} are
significantly lower than their typical values in long-duration GRBs
\citep{Meszaros}. The observed GRB may be just one spike in a longer
and weaker transient below the GBM detection threshold. The weakness
of the burst could be attributed to the extended envelope of the very
massive progenitor star, from which the GRB outflow just barely
managed to escape \citep{Bromberg}.  For this to work, the BH activity
must have persisted for roughly the light crossing time of the star,
$\sim 14(M_\star/100M_\odot)^{1/2}$ s. In particular, the low GRB
luminosity could have resulted from a broader than usual opening angle
of the GRB outflow as it slowed down and widened just before exiting
the stellar envelope. A broad GRB outflow brings the added benefit of
removing the need for a rare alignment between the line-of-sight and
the central axis of the outflow. The parameter fit of LIGO disfavored
orientations where the orbital angular momentum of the BH binary is
misaligned with the line of sight \citep[see Figure 2 in][]{LIGOteam}.

The main advantage of the single star origin for GW150914-GBM is that
it naturally provides a high infall rate of gas around the merging
BHs. The $\gamma$--ray luminosity of GW150914-GBM corresponds to a
mass infall rate of $\sim 1/(\epsilon/10^{-5}) M_\odot~{\rm
  s^{-1}}$, where $\epsilon$ is the efficiency of converting
accreted mass to the observed $\gamma$-rays. An accretion from a
long-lived disk (e.g., originating from the tidal disruption of an
ordinary star) around the BH binary would be typically limited to the
Eddington luminosity \citep{Kamble}, which for a binary mass of $M\sim
65M_\odot$ amounts to $\sim 10^{40}~{\rm erg~s^{-1}}$, a factor of
$\sim 10^9$ lower than the inferred $\gamma$-ray luminosity in
GW150914-GBM.

A future detection of a GRB afterglow would allow to determine the
redshift and precise localization of the GW source \citep[but see the
  upper limits in][]{Smartt,Soares}. Since LIGO detected GW150914 only
shortly after starting to collect data at its improved sensitivity, it
will likely detect many similar events during its future operation. A
population of standard GW sirens with GRB redshifts would provide a
new path for measuring cosmological distances as a function of
redshift to a high precision \citep{Hughes,Nissanke}.

Numerical simulations are required to better characterize the detailed
hydrodynamics and neutrino cooling associated with a binary BH
formation through a dumbbell configuration during the collapse of the
core of a massive star. Magnetic fields could also play an important
role in transporting angular momentum and mediating the collapse of
the two clumps. 

\acknowledgments I thank Peter Edmonds, Dani Maoz, Ramesh Narayan,
Fred Rasio, Martin Rees, Amiel Sternberg and especially the referee
Stan Woosley for insightful comments on the manuscript (with any
remaining errors being all mine), and my family for giving me freedom
to write this paper over a holiday weekend. This work was supported in
part by the NSF grant AST-1312034.

\end{document}